\documentclass[12pt]{iopart}
\usepackage{iopams}  

\usepackage{epsfig}

\begin{document}

\title[DC transport properties and microwave 
absorption in La$_{0.5}$Sr$_{0.5}$CoO$_{3-\delta}$]
{DC transport properties and microwave absorption in bulk ceramic sample
and film of La$_{0.5}$Sr$_{0.5}$CoO$_{3-\delta}$: magnetic
inhomogeneity effects}

\author{B. I. Belevtsev$^{1}$\footnote[1]{To
whom correspondence should be addressed (belevtsev@ilt.kharkov.ua)}, 
N. T. Cherpak$^{2}$,  I. N. Chukanova$^{3}$, \\
A. I. Gubin$^{2}$, V. B. Krasovitsky$^{1}$, A. A. Lavrinovich$^{2}$}

\address{$^{1}$ B. Verkin Institute for Low Temperature 
Physics \& Engineering, National Academy of Sciences, 
Kharkov, 61103, Ukraine}

\address{$^{2}$ A. Usikov Institute for Radiophysics and Electronics, 
National Academy of Sciences, Kharkov, 61085, Ukraine}

\address{$^{3}$ Institute for Single Crystals, National Academy of Sciences, 
Kharkov, 61001, Ukraine}

\begin{abstract}
The DC transport properties and microwave absorption (at 41~GHz) are 
measured in a bulk ceramic sample and a film (220 nm thick) of 
La$_{0.5}$Sr$_{0.5}$CoO$_{3-\delta}$. The bulk sample has been cut from a 
target from which the film was pulsed-laser deposited.  It is found that the 
temperature behavior of DC resistivity and magnetoresistance of the bulk 
sample is quite different from that of the film. This is attributed to 
oxygen depletion of the film as compared with the target. Below Curie 
temperature $T_c$, the film behaves  like a highly inhomogenenous system of 
weak-connected ferromagnetic grains (or clusters).  The microwave study has 
provided further data about inhomogeneity of the samples. It is found that a 
surface layer of the bulk sample has very low conductivity compared with 
the bulk. This can be explained by the oxygen depletion of the surface layer. 
The most important feature of doped cobaltates revealed in this study is the 
following: the microwave conductivity, which should be related mainly to the 
conductivity within the poor-connected grains, increases by the order of the 
magnitude at the transition to the ferromagnetic state. The increase is much 
greater than that of found in the reported DC measurements in doped 
cobaltates of the best crystal perfection. The found microwave effect is 
attributed to inherent magnetically inhomogeneous state of the doped 
cobaltates. 
Relaying on the results obtained it can be suggested that rather low 
magnetoresistance in the doped cobaltates as compared with that of the 
manganites is attributable to their more inhomogenenous magnetic state.  
\end{abstract}

\submitto{\JPCM}

\maketitle

\section{Introduction}
\label{int}
The hole-doped lanthanum cobaltates of the type 
La$_{1-x}$Sr$_{x}$CoO$_{3-\delta}$ ($0 < x \leq 0.5$) with perovskite-related 
structure have  attracted much attention during the last five decades due to 
the range of novel magnetic and transport properties they show (see  
\cite{racca,itoh,ganguly,senaris,yama,mahen,mukher,joy,caciuffo} 
and references therein). The system is important both for fundamental studies 
and for promissing applications \cite{wu}. The undoped cobaltate LaCoO$_3$ 
($x=0$) is an insulator. In the intermediate range of Sr doping 
($0.3 < x < 0.5$) it is a high conductive  metal, which shows ferromagnetism 
(FM) below Curie temperature $T_c = 240 \pm 10$~K 
\cite{itoh,ganguly,senaris,mahen}. An interest in the doped 
cobaltates has quickened in recent years after discovery of, so called, 
colossal magnetoresistance (CMR) in the related FM perovskite oxides, 
named doped manganites, of the type 
La$_{1-x}$A$_x$MnO$_3$ (where $A$ is a divalent alkaline-earth element 
like Ca, Sr, Ba) (see reviews \cite{kaplan,dagotto,nagaev} 
as an introduction to the problem). In the doped manganite films, the 
magnetoresistance (MR), defined as $\delta (H)=[R(0)-R(H)]/R(0)$, was found 
to be more than 90\% at fields $H$ about 60 kOe in the neighbourhood of
room temperature. This has triggered intensive theoretical and 
experimental studies by a large number of scientific groups around 
the world. In spite of this, a clear understanding of the CMR is not yet 
available. In the Sr doped cobaltates, the MR magnitude is found to be much 
less (only a few percents). It is believed that elucidation of reasons of 
this large difference in MR behavior of manganites and cobaltates can be 
helpful for understanding of CMR nature.
\par
It follows from the known studies \cite{dagotto,nagaev} that doped 
manganites are always magnetically inhomogoneous to some extent. The degree
of inhomogeneity depends on the doping level and preparation conditions. 
More exactly, two main types of inhomogeneity sources in the doped manganites 
are distinguished at present: extrinsic and intrinsic. Extrinsic sources are 
due to various technological factors of sample preparation.  They can cause 
chemical-composition inhomogeneity (for example, in an oxygen concentration), 
structural inhomogeneity (polycrystalline or granular structure) and others. 
Intrinsic ones are believed to be due to thermodynamical reasons and can 
lead to phase separation on two phases with different concentration of the 
charge carriers  \cite{dagotto,nagaev}. The inhomogeneity effect on magnetic
and transport properties of manganites is usually minimal at the optimal 
doping ($x \approx 0.33$) 
but can be significant at low doping level. Needless to say that any of the 
above-mentioned types of inhomogeneity is associated with perturbation of 
the magnetic order and, therefore, with magnetic inhomogeneity.
\par
The types of inhomogeneity, outlined for manganites, reveal themselves in 
doped cobaltates as well \cite{itoh,ganguly,senaris,mukher,joy,caciuffo}. 
What's more the cobaltates appear to be more prone to phase separation 
than doped manganites. The system 
La$_{1-x}$Sr$_{x}$CoO$_{3-\delta}$ with $0.18 \leq x \leq 0.5$ shows 
evidence of phase separation. Beginning with low dopant concentration 
($x > 0.1$) an inhomogeneous distribution of the Sr$^{2+}$ ions takes place.
This results in the segregation of the material into hole-rich FM regions 
and a hole-poor semiconducting matrix. It was suggested \cite{itoh} that this 
inhomogeneous state should appear as, so called, magnetic cluster-glass 
phase. The cluster glass is some set of clusters, formed due to short-range 
FM ordering at the Curie temperature $T_c$. The clusters are embedded in a 
non-ferromagnetic matrix. Below $T_c$, cluster-glass system is expected to 
demonstrate spin-glass behavior for temperature decreasing. This suggestion 
was justified to some degree by AC susceptibility 
measurements \cite{itoh,mukher}. For this reason some of the scientific 
groups believed that cluster-glass state persists up to concentration 
$x = 0.5$ \cite{itoh,mukher}. By this is meant that the clusters remain 
isolated from one another up to this concentration and, therefore, this 
state does not have a long-range FM order. It is known, however, that volume 
fraction of  the FM regions increases with the Sr 
concentration \cite{senaris,caciuffo}. 
At $x > 0.25$ the metallic FM regions percolate magnetically as well as 
conductively \cite{caciuffo}, that is a long-range FM order does occur in this
system at high enough Sr concentration. The authors of references 
\cite{ganguly,joy} have arrived to the same conclusion. It should be 
taken into account, however, that inpenetrating hole-poor matrix and some 
isolated clusters in it persist to $x=0.50$ \cite{senaris}. 
\par
From the aforesaid, it might be assumed that compound 
La$_{1-x}$Sr$_{x}$CoO$_{3-\delta}$ is magnetically inhomogeneous in the 
whole range $0.18 \leq x \leq 0.5$, where FM ordering manifests itself. 
This can lead to peculiar magnetic and transport properties. In studies of 
transport properties of FM perovskites, it is most common to use of direct 
current (DC) measurements. For doped manganites, however, the microwave 
studies were undertaken in several groups \cite{lofland,yates,solin}. 
These 
have brought some interesting results and, in particular, have revealed that 
wicrowave conductivity is more sensitive  than DC conductivity to the 
changes in magnetic state of 
manganites at the paramagnetic-ferromagnetic transition. In this way the
microwave data disclose magnetic inhomogeneity in manganites. It can be 
expected that in the case of doped cobaltates the microwave study can bring 
some interesting data about their magnetic inhomogeneity and its sources as 
well.  Some previous microwave studies 
of cobaltates have not come to our knowledge. All this motivate us to study 
transport properties in La$_{0.5}$Sr$_{0.5}$CoO$_{3-\delta}$ not only 
through the use of DC measurements, but by measurements of microwave 
absorption as well.  A choice of composition ($x=0.5$) was determined to 
some extent by the fact that this compound is expected to find the most 
application in advanced technology~\cite{wu}.
\par 
A bulk ceramic sample and a film (prepared by pulsed-laser deposition) were 
studied.
We have found that DC transport characteristics (magnitude and temperature 
behavior of the resistivity and these of the MR) are distinctly affected by 
magnetic inhomogeneity of the samples below Curie temperature $T_c$. 
On the strength of these data it is concluded that the film studied is an 
inhomogeneous system of weak-connected FM grains (or clusters) at $T<T_c$.
The microwave absorption study has provided further data about inhomogeneity 
of the samples. In particular, it is found that a surface layer of the bulk 
sample has very low conductivity comparing with the bulk. This is explained 
by the surface oxygen depletion of the bulk sample. The DC method of
transport measurements is not sensitive to this type of inhomogeneity when 
the oxygen-depleted layer is fairly thin.
\par
The most important feature of doped cobaltates revealed in this study is the 
following: the microwave conductivity, which should be related mainly to the 
conductivity within the poor-connected grains, increases by the order of the 
magnitude at the transition to the FM state. The increase is much greater 
than that of found in the reported DC measurements in doped cobaltates of 
best crystal perfection, that is with the minimal effect of extrinsic 
sources of an inhomogeneity. Thus the found microwave effect can be safely 
attributed to inherent magnetically inhomogeneous state of doped cobaltates. 
Relaying on the results obtained it can be suggested that rather low MR in 
the doped cobaltates as compared with that of the CMR manganites is 
attributable to their more inhomogenenous magnetic state. 

\section{Experiment}
The samples studied are the pulsed-laser deposited (PLD) film and the bulk 
sample (cut from a target from which the film was deposited). The disk-shaped 
ceramic target with a composition La$_{0.5}$Sr$_{0.5}$CoO$_{3}$ was prepared 
by a standard solid state reaction technique. The final annealing was done 
at temperatures 1000$^{\circ}$C and 1200$^{\circ}$C. X-ray diffraction does 
not reveal any inclusions of unreacted components in the target, suggesting 
that it was homogeneous in chemical composition. The target was 
polycrystalline with rather large grain size (in the range 40-70 $\mu$m). 
It was porous (as it usually is for this type of samples) with void 
content about 45\%. The lattice parameter for the target indexed for a 
pseudo-cubic perovskite-like cell is found to be $a_p = 0.383\pm 0.001$~nm, 
that agrees closely with measurements of reference \cite{ganguly}. 
\par
The film ($220\pm 20$~nm thick) was grown  on (001) oriented LaAlO$_3$ 
substrate. A PLD system with Nd-YAG laser operating at 1.06 $\mu$m was used 
to ablate the target. The pulse energy was about 0.39 J with a repetition 
rate of 12 Hz and pulse duration of 10 ns. A standard substrate arrangement 
normal to the laser plume axis was used. The other details of 
the PLD technique employed were described previously \cite{usoskin}. 
The film was deposited with a substrate temperature 880$\pm 5^{\circ}$C in 
an oxygen atmosphere at pressure about 8 Pa. The oxygen pressure in the
PLD chamber was increased up to about $10^{5}$ Pa after deposition, the film 
was cooled down to room temperature in this oxygen atmosphere.
\par
The resistance of the samples as a function of temperature and magnetic field 
$H$ (up to 20 kOe) was measured using a standard four-point probe technique. 
The available cryostat with a rotating electromagnet makes it possible to
measure DC resistance with different directions of $H$ relative to the
plane of the film. 
\par 
Microwave conductivity of the samples was determined at frequency 
$\nu = 41$~GHz from measurements of the reflection coefficient $R_r$ for 
samples placed in a waveguide with $5.2\times 2.6$ mm$^2$ cross-section. The 
samples were inclined  at a $10^{\circ}$ angle to the waveguide broad wall 
(an angle between  the normal to the sample and longitudinal waveguide axis 
is $\theta = 80^{\circ}$ in the plane of $\vec{E}$, where $\vec{E}$ is the 
vector of microwave electric field in the waveguide) \cite{cherpak1}. The technique 
employed similar to that 
used in IR experiments \cite{somal}, in which a $p$-polarized beam has been 
incident on a specimen at a grazing angle. Such approach allows to enhance 
the sensitivity of reflection measurements. It was shown previously 
\cite{cherpak2} that with a knowledge of complex refraction coefficients of 
a film, $n = \sqrt{\epsilon\mu}\,$, and a substrate,
$n_s = \sqrt{\epsilon_{s}\mu_{s}}\,$ (where $\epsilon$, $\mu$ and 
$\epsilon_{s}$, $\mu_{s}$ are complex dielectric permittivity and magnetic 
permeability of the film and the substrate, respectively), and their 
thicknesses $d$ and $d_s$, it is possible to calculate reflectivity 
coefficient for the plane wave. We have used relations 
$\epsilon = 1+\sigma_{mw}/i\omega\epsilon_{0}$ (where $\sigma_{mw}$ is 
the microwave conductivity) for the films and $\mu_{s}=1$ for
dielectric substrate. Having experimental data for reflection 
coefficient $R_r$, $\epsilon_{s}$, $d$ and $d_s$, it is possible 
to calculate with help of backward transformation the values of
$\sigma_{mw}$ and corresponding resistivity $\rho_{mw}= 1/\sigma_{mw}$. 
At micromave studies the microwave surface resistance 
$R_s=(\omega \mu\mu_{0}/2\sigma_{mw})^{1/2}$ is often considered as an 
important parameter. This can be calulated, based on $\sigma_{mw}$ data
for appropriate values of $\mu$.  
\par
The reflectivity as a function of temperature has been measured using a 
microwave phase bridge in the 6 mm wavelength range. In practice, change of 
the signal microwave power reflected from the samples was measured. This 
change contains information about the sample conductivity variation. A 
calibration procedure (which was done by means of the high-$T_c$ 
superconducting YBaCuO film at $T=77$~K) allows us to determine the absolute 
values of the microwave conductivity in the samples studied with an accuracy 
which is believed to be within range $-10\%$ to $+100\%$. An ``asymmetry'' 
of maximal probable error results from non-linear dependence of $R_r$ on 
$\sigma_{mw}$. The calculated absolute values of  $\sigma_{mw}$ (or 
$\rho_{mw}$) depend essentially on the used value of permeability $\mu$. We 
have taken $\mu=1$ at such calculations. A discussion of appropriateness of 
this choice is presented in the next section of this paper. On the other 
hand, the
relative temperature changes in $R_r$ were measured with higher precision,
which results in an accuracy of relative temperature changes in 
$\sigma_{mw}$ about $\pm 4\%$ for the bulk sample and $\pm 10\%$ for the 
film. 

\section{Results and discussion}
\subsection{DC transport properties}
The temperature dependence of DC resistivity, $\rho (T)$, of the bulk 
ceramic sample is presented in figure 1a. The dependence agrees well with 
the known results  on La$_{1-x}$Sr$_{x}$CoO$_{3-\delta}$ for 
$0.3 \leq x \leq 0.5$ \cite{ganguly,senaris,yama}. The resistivity values 
correspond to the known data on the ceramic doped cobaltates as 
well \cite{ganguly}. The $\rho$ values presented in figure~1a were 
calculated taking into account the porosity of the 
sample\footnote[2]{This was done in the simplest way using the fractional 
part (55\%) of actual cross-section of the sample at calculation of
the sample resistivity. In the case of fairly homogenenous pore distribution 
(as shown by SEM study of the sample) the error in calculated $\rho$ 
values should not exceed about 10\%.}, 
so that they  present, 
within certain limits, the resistivity of compact material. In 
La$_{0.5}$Sr$_{0.5}$CoO$_{3-\delta}$ with fairly perfect crystalline 
structure and $\delta$ close to zero, the resistivity values are reported to 
be about 100 $\mu \Omega$~cm (or even less) at the room temperature
\cite{racca,senaris,madhukar}. 
The higher value of $\rho$ in the 
sample studied is to be attributed to its polycrystalline structure and some
oxygen defficiency. The ratio of resistances at 300~K and 4.2 K, is more 
than 2, that corresponds to the data of 
references \cite{senaris,mahen} for the samples with fairly good 
crystal quality. The $T_c$ value for the sample studied was found to be 
about 250~K by AC susceptibility measurements \cite{beznosov}. 
\par 
It is seen in figure~1a that the $\rho (T)$ behavior is metallic 
($d\rho /dT > 0$) in the whole temperature range investigated, below and 
above $T_c$.  The $\rho (T)$ dependence exhibits a change of slope at $T_c$ 
as a result of spin disorder scattering. This is reflected by the 
temperature behavior of $d\rho /dT$ (figure~1a) which has a maximum near 
$T_c = 250$~K. Such behavior of $\rho (T)$ is typical in FM 
metals \cite{vonsov}. The point is that the resistivity of FM metal has a 
quite pronounced contribution from the electron scattering on spin disorder 
(apart from the usual contributions from crystal-lattice defects and 
electron-phonon scattering) \cite{vonsov}. This contribution represents, so 
called, magnetic part of the resistivity, $\rho_{m}(T,H)$, which depends on 
the magnetization. With a rise of magnetization at the transition to the FM 
state, $\rho_{m}$ drops appreciably, which is a reason for an enhanced 
resistivity decrease  below $T_c$, when going from high to lower temperaure. 
This takes place in the sample studied as well (figure~1a). The external 
magnetic field enhances the spin order, that leads to a decrease in the 
resistivity. That is why the FM metals are characterized by a negative MR. 
The negative MR of the ceramic sample is found to have a maximum absolute 
value at  
$T=T_c$ [$\delta (H)= 2.15$~\% at $H = 20$~kOe]. It goes down rather steeply 
to $\delta (H) \ll 1\%$ for temperature deviating to either side from $T_c$. 
Such temperature behavior of MR is expected for the FM metals of fairly good 
crystal perfection. 
\par
We turn now to the DC conductivity behavior of the PLD film studied 
(figure~2a). 
This is quite different from that of the bulk sample. First, the resistivity 
is higher than in the bulk sample (compare with figure~1a). Second, 
$\rho (T)$ curve has a maximum at $T_{p} \approx 250$~K  and a minimum at 
$T_{m} \approx 107$~K. A decrease in resistance below $T_c$ is rather small 
(by about 10 \% from the maximum value). Third, the MR of the film is found to 
be less than in the bulk sample (a  maximum value of $\delta(H)$ is about
0.8 \% at $T\approx 230$~K and  $H=20$~kOe, as shown in figure~3).
\par
Consider now the temperature behavior of the resistivity of the film more 
closely. The first thing that catches the eye is that for $T> T_{p} = 250$~K 
and for $T < T_{m} = 107$~K the temperature behavior is non-metallic 
($dR/dT < 0$).  This type of $\rho (T)$ behavior (that is a maximum near 
$T_c$ and a minimum about 100 K) is typical for cobaltates 
La$_{1-x}$Sr$_{x}$CoO$_{3-\delta}$ with low doping level 
$0.2 \leq x \leq 0.3$ \cite{senaris}, that is for low charge-carrier 
concentration. The appearence of this type of dependence in PLD film with a 
nominal composition La$_{0.5}$Sr$_{0.5}$CoO$_{3-\delta}$ should be 
attributed to some considerable oxygen deficiency (large value of $\delta$), 
that can cause the significant decrease in the charge-carrier 
concentration \cite{madhukar,liu}. Beside reducing the number of carriers,
the oxygen deficiency is connected with the presence of oxygen vacancies 
sites, which hinder the carrier motion. It was shown in studies
\cite{madhukar,liu} that for sufficiently high values of 
$\delta$ the La$_{0.5}$Sr$_{0.5}$CoO$_{3-\delta}$ films can be even 
insulating. 
\par
It should be mentioned, that $\rho (T)$ dependence, shown in figure~2a 
(that is a maximum at $T=T_p$ and a minimum at low temperature) is typical 
for systems of FM grains (or clusters) with rather weak interconnection.
For example, similar dependences were found in polycrystalline 
manganites \cite{mahen2,andres,rozen}. For these compounds, the paramagnetic
phase is non-metallic, and for this reason the temperature dependence of 
resistance has a peak at a temperature $T_p$ at which an infinite 
percolating ferromagnetic cluster is formed at the transition.  The grain 
boundaries in FM perovskite oxides are usually poorly conductive or can be
even dielectric. Therefore, in the limiting case, a polycrystalline sample is
just a granular metal, that is a system of metal grains embedded in an 
insulator. The conductivity of granular metals is determined by the 
tunnelling of 
charge carriers through the boundaries between the grains. The tunnelling can 
be activated or non-activated, depending on the thickness of boundaries and 
the temperature. Real granular systems are inhomogeneous in strength of 
tunnelling barriers between grains and, therefore, in the probability of 
charge-carrier tunneling. These systems are actually percolating and their 
conductivity is determined by both intragrain and intergrain transport 
properties.
\par
The conductivity of granular metal is conditioned by two processes: 
tunnelling and thermal activation of the charge carriers. If the activation 
energy $E_a$ is rather 
low, the conductivity at high enough temperature, $kT>E_a$, is non-activated. 
In this case the system behaves as a (``bad'') metal with a positive 
temperature  coefficient of resistivity ($dR/dT > 0$). Since granular metals 
are percolating systems, their conductivity is determined by the presence 
of ``optimal'' chains of grains with maximal probability of tunnelling for 
adjacent pairs of grains forming the chain. These``optimal'' chains have 
some weak high-resistive links. At low enough temperature the relation 
$kT<E_a$ can become true for these links, and, hence, the measured DC
conductivity of the system will become activated. 
\par
We have considered here the most obvious reasons for appearence of the 
resistance minimum at low temperature in inhomogeneous FM systems. Some more 
specific models of this phenomenon in doped manganites are presented in 
references \cite{andres,rozen}. The above-mentioned general reasons for the 
resistance minimum can be applied to the film studied as well. On the basis 
of our gained experience, we can safely assume that PLD cobaltate films 
obtained in the above-described conditions are polycrystalline. There is a
reason to believe that distribution of oxygen vacancies in perovskite oxides 
is inhomogeneous. The vacancies are more likely to reside at the grain 
boundaries \cite{yates,hossain}, that increases the structural and magnetic 
inhomogeneity. Beside this, the doped cobaltates are believed to have an 
intrinsic source of 
inhomogeneity: the phase separation, that is a segregation of material into 
FM metallic clusters embedded in a hole-poor semiconducting matrix 
\cite{itoh,senaris,mukher,caciuffo} (see discussion in Sec.~\ref{int}). 
In cobaltates with low charge-carrier concentration, the FM clusters are 
weakly connected or even isolated, so that the system can behave as a 
granular FM metal. The $R(T)$ dependences quite similar to figure~2a were 
observed previously  in 
La$_{1-x}$Sr$_{x}$CoO$_{3-\delta}$ with $0.2 \leq x \leq 0.3$ \cite{senaris}. 
The authors of reference \cite{senaris} speculate that for this composition 
the metallic conduction is only established within interval 100 K 
$ < T < T_c$, but below 100 K the increasing population of low-spin Co(III) 
ions in the matrix reestablishes a non-metallic temperature dependence. May 
be this is a reason for the observed $\rho (T)$ behavior in the film studied. 
In any case, however, it is evident from figure~2a that the clusters (or FM 
grains) in the film have a weak connectivity.
For this reason the resistance shows only slight (10~\%) decrease at the 
transition to FM state. This change is far less than that in the bulk
sample, which shows a twofold decrease in resistivity (figure~1a).  
\par
From the aforesaid, it appears that the film studied behaves as a 
system of weakly connected FM grains (or clusters). The MR data obtained 
support this viewpoint. First, a maximum value of the MR in the 
film near the temperature of transition to the FM state is less than that of 
in the bulk sample. This is quite reasonable, since the measured MR of a 
granular FM system near $T_c$ should be less that an intrinsic intragrain MR 
in the cases that grains have poor interconnectivity. This obvious effect has 
been seen previously, for example, in polycrystalline films of doped 
manganite \cite{rivas}. Second, in polycrystalline samples the significant 
contribution to MR comes from grain boundaries, and this contribution 
increases with decreasing temperature. Discussion of possible mechanisms of 
this extrinsic type of MR can be found in references 
\cite{gupta,hwang,evetts}.
The film studied shows indeed a continuous increase in MR for decreasing 
temperature (for the temperatures fairly below $T_c$) (figure~3). This is an 
expected behavior for polycrystalline samples with poor enough intergrain
connectivity \cite{gupta,hwang}. In contrast with it, for  cobaltate and 
manganite samples with fairly good crystal perfection or even for 
polycrystalline samples of these materials but with a good intergrain 
connectivity the MR has a peak near $T_c$ and goes nearly to zero with 
decreasing temperature \cite{yama,mahen,gupta}. This effect has found a 
verification in this study as well for the bulk sample (see above). In 
summary, the DC resistance and MR of the 
 film show that it makes a system of weakly connected grains (or clusters). 
This inhomogeneous phase state in the (nominally) high-doped cobaltate 
($x=0.5$) should be attributed mainly to rather significant oxygen deficiency.
\par
At the end of this subsection it should be added  that the MR of the film is 
strongly anisotropic (figure~3). Although (as far as we know) nobody else 
has yet reported this effect in doped cobaltate films, the anisotropic MR is 
not  surprising for PLD  films of FM perovskite oxides \cite{boris2}. In the 
present study it is found that the absolute values of the negative MR in 
fields parallel to the film plane are much above that of in the 
perpendicular fields (figure~3). Since conductivity of doped cobaltates 
increases with an enhancement of the magnetic order, this behavior just 
reflects the point 
that the magnetization increases more easily in a magnetic field parallel to 
the film plane. This MR anisotropy is connected with the FM state. For this
reason it disappears for $T > T_c$. This effect (which is not a main goal
of this paper) is to be discussed in more detail elsewhere.

\subsection{Microwave absorption measurements and comparison with DC 
transport properties}
As stated above, the $\rho_{mw}$ values, obtained in the present microwave
study, are deduced taking $\mu=1$. Although some direct measurements of $\mu$ 
in La$_{1-x}$Sr$_{x}$CoO$_{3}$ system for $0.2 \leq x \leq 0.5$ have not 
come to our knowledge, this choice seems to us as the most appropriate and
credible. In FM metals  $\mu$ is equal to unity above Curie temperature. 
Below $T_c$ it can be considerably larger than unity, but only for
rather low frequencies. For frequency decreasing the value of $\mu$ 
decreases, so that at high enough frequency it is close to unity 
\cite{vonsov}. This general for FM metals tendency is found to be true 
for related perovskite-like oxides as well. For example, in polycrystalline 
La$_{0.7}$Sr$_{0.3}$MnO$_{3}$ the magnitude of $\mu$ decreases significantly
for frequency above 1 MHz, so that at $\nu=100$~MHz the magnitude of $\mu$ is 
only moderately greater than unity \cite{wang}. It can be expected that 
at far  higher frequency $\nu=41$~GHz, used in this study, $\mu$ should be 
quite close to unity. Some additional support to this can be seen in the
fact, that  the best agreement between the DC and microwave resistivities in 
the film studied for $T > T_c$ has been found  at $\mu=1$ (see below).  For 
$\mu >1$
an agreement is much worse. In any case, even if $\mu$ is somewhat larger 
than unity in the samples studied it cannot effect essentially the main 
findings and conclusions of this paper outlined below. 
\par
A comparison of the temperature dependences of $\rho_{mw}$ and DC resistivity 
($\rho$) for the bulk sample and the film can be made with figures~1 and 2. 
Let us begin with the bulk sample (figure~1). Its $\rho_{mw}(T)$ dependence is 
non-monotonic with a maximum at $T_{p} \approx 220$~K, so that  the 
corresponding $R(T)$ behavior is non-metallic at $T> T_p$. This  is in  
sharp contrast to the $\rho(T)$ dependence of the bulk sample (figure~1a). 
As discussed above, the non-monotonic $R(T)$ dependences with a maximum 
below $T_c$ are inherent in doped cobaltates with low density of the charge 
carriers and thus with the high resistivity. Indeed, the $\rho_{mw}$ values, 
estimated for $\mu=1$, are found to be more than two orders of the magnitude 
greater than corresponding DC values of 
the resistivity, indicated in figure~1a. The skin depth in the bulk sample  
is about 80~$\mu$m at the used frequency. Therefore, the found $\rho_{mw}(T)$ 
behaviour is to be attributed to the properties of a rather thin surface 
layer of the bulk sample, which is poor of the charge carriers, most likely, 
due to an oxygen depletion. It is known that ceramic samples of 
La$_{0.5}$Sr$_{0.5}$CoO$_{3-\delta}$ are always oxygen deficient with 
$\delta$ up to 0.06 \cite{senaris,caciuffo,kumar}. But even greater oxygen 
deficiency 
can be expected at the surface of the samples comparing with the bulk. 
This suggestion has been made previously in reference \cite{kumar} 
based on the specific features in the temperature curves 
of the AC susceptibility of ceramic La$_{0.5}$Sr$_{0.5}$CoO$_{3-\delta}$. 
The microwave study in this paper supports this suggestion. 
\par
Contrary to the bulk sample, where the microwave absorption is confined 
solely to the surface skin-depth layer, the temperature behavior of 
$\rho_{mw}$ in the film (figure~2b) is determined by response of the whole 
film volume to the microwave action, since the skin depth is far larger than 
the film thickness.  The maximal values of $\rho_{mw}$ in the film for 
$T>T_c$ (where $\mu$ is sure equal to unity) are found to be quite close to  
the film DC resistivity (figure~2)\footnote[3]{It should be noted that the
main uncertainty in the DC resistivity of the film studied is determined by
an accuracy in the film thickness estimation. Taking this into account, the
expected accuracy in the film DC resistivity is about $\pm 10 \%$.}. 
The $\rho_{mw}(T)$ dependence has a 
peak at the nearly same temperature $T_p$ as the $\rho (T)$ curve, but it 
does not show a pronounced minimum at $T_{m} = 107$~K as the $\rho(T)$ 
curve does (figure~2). The most sharp distinction between the temperature 
behaviour of the microwave and DC resistivities, found in this study, is the 
following. It is seen in figure~2b that for temperature decreasing below 
$T_c$ the $\rho_{mw}$ magnitude decreases by a factor about ten.
This change is really huge in comparison with the 10~\% decrease in the film 
DC resistivity in the same temperature range (figure~2a). 
\par
The greater changes in $\rho_{mw}$ at the paramagnetic-ferromagnetic 
transition when compared with that of in the DC resistivity  have been seen 
previously in the microwave \cite{lofland,solin,lofland2} and IR 
optical \cite{kim} studies  of polycrystalline doped manganites. It is 
apparent that this effect is typical of inhomogenenous perovskite oxides for 
electromagnetic waves of not too high frequency. In some cases 
\cite{solin,kim}  the 
high-frequency resistivity is found to be far less than the DC resistivity 
in the whole temperature range studied (below and above $T_c$). In other 
cases \cite{lofland,lofland2} the microwave resistivity is found to be nearly 
equal to the DC resistivity (and their temperature behaviour with a maximum 
at $T=T_p$ are found to be exactly the same) in the temperature range above 
and near $T_c$ (or $T_p$), and only below $T_p$ the microwave resistivity
becomes far less than the DC one. The latter type of behaviour is found in 
this study of doped cobaltate La$_{0.5}$Sr$_{0.5}$CoO$_{3-\delta}$.  
\par
The reported results of microwave and optical studies of doped 
manganites \cite{lofland,solin,lofland2,kim} appear, at first sight, as 
inconsistent and bizarre from standpoint of Drude theory of optical 
properties of metals \cite{ziman,noskov}. According to that, for 
electromagnetic waves with low enough frequency ($\omega \ll 1/\tau$, where 
$\tau$ is the electron relaxation time in a conductor) the optical 
conductivity should not depend on the frequency and be equal to the DC 
conductivity. It is true, however, only for homogeneous systems. Consider 
this question in more detail. In doing so we shall use the more general term 
``optical conductivity'' implying that this has the same meaning as the 
term ``microwave conductivity'' for low enough frequency.  The point is that 
the DC and optical conductivities are determined by quite different physical 
processes. The DC conductivity is defined by the ability of the charge 
carriers to propagate through a conducting system. It is apparent that this 
ability is affected profoundly by any kind of structural or phase 
inhomogeneity, especially when the system is just some disordered mixture of 
metal and insulator. Take again as an example a granular FM metal with poor 
intergrain connectivity. At the transition to the FM state the 
intragrain conductivity of this inhomogenenous system can increase profoundly, 
but the measured DC conductivity of the whole sample can show only rather 
weak increasing due to the weak intergrain connectivity (see above an 
extended discussion of this matter). For this reason the absolute values and 
temperature dependences of the DC resistivity and MR in FM perovskite oxides 
depend crucially on preparation conditions, which determine the degree of 
inhomogeneity of samples. In addition, there can be the intrinsic source of 
inhomogeneity (the phase separation) which shows up even in single-crystal
samples. And this source, as it appears from the reported studies, is more 
important in doped cobaltates than in manganites. 
\par
The optical conductivity is determined by the ability of metal to absorb the 
energy of the electromagnetic field. For low frequency 
$\omega < kT$ (microwave and IR range) the main contribution to it comes from
classical absorption, at which the charge carriers are accelerated by the
electric field of electromagnetic wave \cite{ziman,noskov}. The absorption 
ability is characterized by the processes of transformation of energy flux of 
electromagnetic field to a thermal flux through the scattering of the charge
carriers on phonons, impurities and other lattice defects \cite{noskov}. In 
FM metals the electron scattering on the spin disorder can give an 
appreciable contribution to these processes. 
\par
Taking into account the DC transport properties of the film studied and 
results of the previous studies of La$_{0.5}$Sr$_{0.5}$CoO$_{3-\delta}$ 
\cite{itoh,ganguly,senaris,mukher,joy,caciuffo}, it can be safely assumed 
that the 
film is oxygen-depleted and inhomogeneous. With approaching and 
crossing $T_c$ when going from high to lower temperature the film transforms 
into magnetically inhomogeneous state of weakly connected FM clusters (or 
grains) together with some isolated FM clusters and an inpenetrating 
hole-poor insulating matrix. It is clear that the main contribution to the 
changes in light (or microwave) absorption and thus in the optical 
conductivity at this transition will be determined by formation of metallic 
FM regions even when they are not or poor connected to one another. In this
way the changes in the optical absorption in doped cobaltates at the 
transition to the FM state reflect rather closely the corresponding changes 
in the intragrain (or intracluster) conductivity. In this study the 
approximately tenfold increase in the microwave conductivity is found 
at this transition, although the reported DC measurements of doped cobaltates 
have shown about threefold increase in the DC conductivity at 
most \cite{senaris,yama,mahen}. This difference in the DC behavior is quite 
reasonable, since even cobaltate samples of the best crystal perfection 
are magnetically inhomogeneous owing to the above-mentioned inherent sources
of inhomogeneity.
\par
In summary, the changes in microwave conductivity at the transition to the FM 
state reflect more closely the real changes in intragrain conductivity of 
doped inhomogeneous perovskite oxides than their DC conductivity. At first 
glance the large (by the order of the magnitude) increase in the microwave 
conductivity at the transition to the FM state which is found in this study 
in La$_{0.5}$Sr$_{0.5}$CoO$_{3-\delta}$ film should involve the 
corresponding high magnitude of the MR. This is not the case, however,
since, as was mentioned above, the MR in inhomogeneous granular or cluster
system, obtained by DC measurements, is always less than the actual 
intragrain (or intracluster) MR. This can be a plausible reason for the 
much lower MR magnitude in cobaltates in comparison with that of in CMR 
doped manganites. The known studies testify that doped cobaltates are prone 
to different types of inhomogeneity to a much greater extent than 
manganites.

\section*{Conclusions} 
We have studied DC transport properties and microwave absorption in 
the bulk ceramic sample and the film of La$_{0.5}$Sr$_{0.5}$CoO$_{3-\delta}$. 
It is found that DC transport characteristics (magnitude and temperature 
behavior of the resistivity and these of the MR) are distinctly affected by 
magnetic inhomogeneity of the samples below Curie temperature $T_c$. Under 
gaining enough experimental and theoretical data on this matter, it is 
possible to apply the DC transport measurements to reveal a structural 
and magnetic inhomogeneity in FM perovskite oxides. Among other things, 
analysis of the DC transport data can be used for an assessment of the 
degree of inhomogeneity and for identification of specific types of 
inhomogeneities (such as the granular structure). It follows clearly from our 
DC measurements that the film studied is an inhomogeneous system of 
weakly connected FM grains (or clusters) at $T<T_c$.
\par
On the other hand, however, a single method should not be enough for 
characterization of the inhomogeneity in FM perovskite oxides.  The study of 
the microwave absorption has provided an expected support for this apparent 
truth. It is found, in particular, that a surface layer of the bulk 
sample has very low conductivity comparing with the bulk. This can be 
explained by the oxygen depletion of the surface layer. The DC method of
transport measurements is not sensitive enough to this type of inhomogeneity 
when the oxygen-depleted layer is fairly thin. The most important feature of 
doped cobaltates revealed in this study is the following: the 
microwave conductivity, which should be related mainly to the conductivity 
within the poor-connected grains, increases by the order of the magnitude at
the transition to the FM state. The increase is much greater than that of
found in the reported DC measurements in doped cobaltates of the best crystal 
perfection, that is with the minimal effect of extrinsic sources of an 
inhomogeneity. Thus the found microwave effect can be attributed to 
magnetically inhomogeneous state which is believed to be inherent in the 
doped cobaltates 
for any doping level. Relaying on the results obtained it can be concluded 
that significantly lower MR in the doped cobaltates as compared with that of 
the CMR manganites can be attributed to their more inhomogenenous magnetic 
state. 


\newpage
\section*{Figures}

\begin{figure}
\caption{Temperature dependences of the resistivity, $\rho$, and its 
derivative, $d\rho /dT$, from DC measurements (a), and the microwave  
resistivity, $\rho_{mw}$, from the microwave measurements at 
41 GHz (b) for the bulk sample of La$_{0.5}$Sr$_{0.5}$CoO$_{3-\delta}$.
The values of $\rho_{mw}$ are deduced taking $\mu=1$.} 
\label{Fig.1}
\end{figure}

\begin{figure}
\caption{Temperature dependences of the resistivity, $\rho$, from  DC 
measurements (a) and the microwave resistance, $\rho_{mw}$, from the 
microwave measurements at 41 GHz (b) for the PLD 
La$_{0.5}$Sr$_{0.5}$CoO$_{3-\delta}$ film. The values of $\rho_{mw}$ are 
deduced taking $\mu=1$.} 
\label{Fig.2}
\end{figure}

\begin{figure}
\caption{Magnetoresistance $\delta (H)=[R(0)-R(H)]/R(0)$ at $H=20$~kOe for 
the fields $H_{\parallel}$ and $H_{\perp}$, applied parallel and 
perpendicular to the film plane. In both cases the fields were perpendicular 
to the transport current. The solid line presents a B-spline fitting.}
\label{Fig.3}
\end{figure}

\begin{figure}
\centerline{\epsfig{file=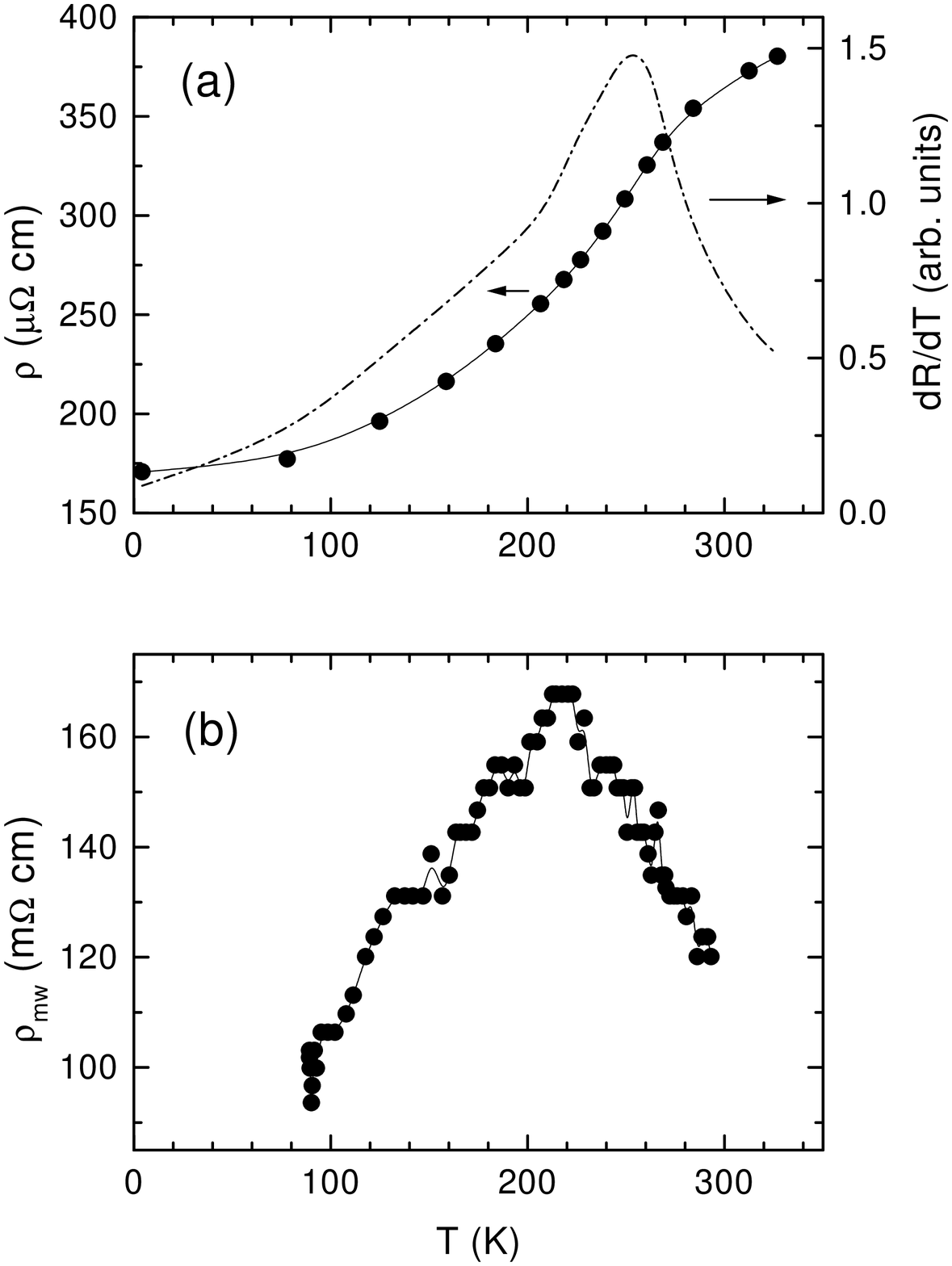,width=12cm}}
\vspace{15pt}
\centerline{Fig.1 to paper Belevtsev et al.}
\end{figure}

\begin{figure}
\centerline{\epsfig{file=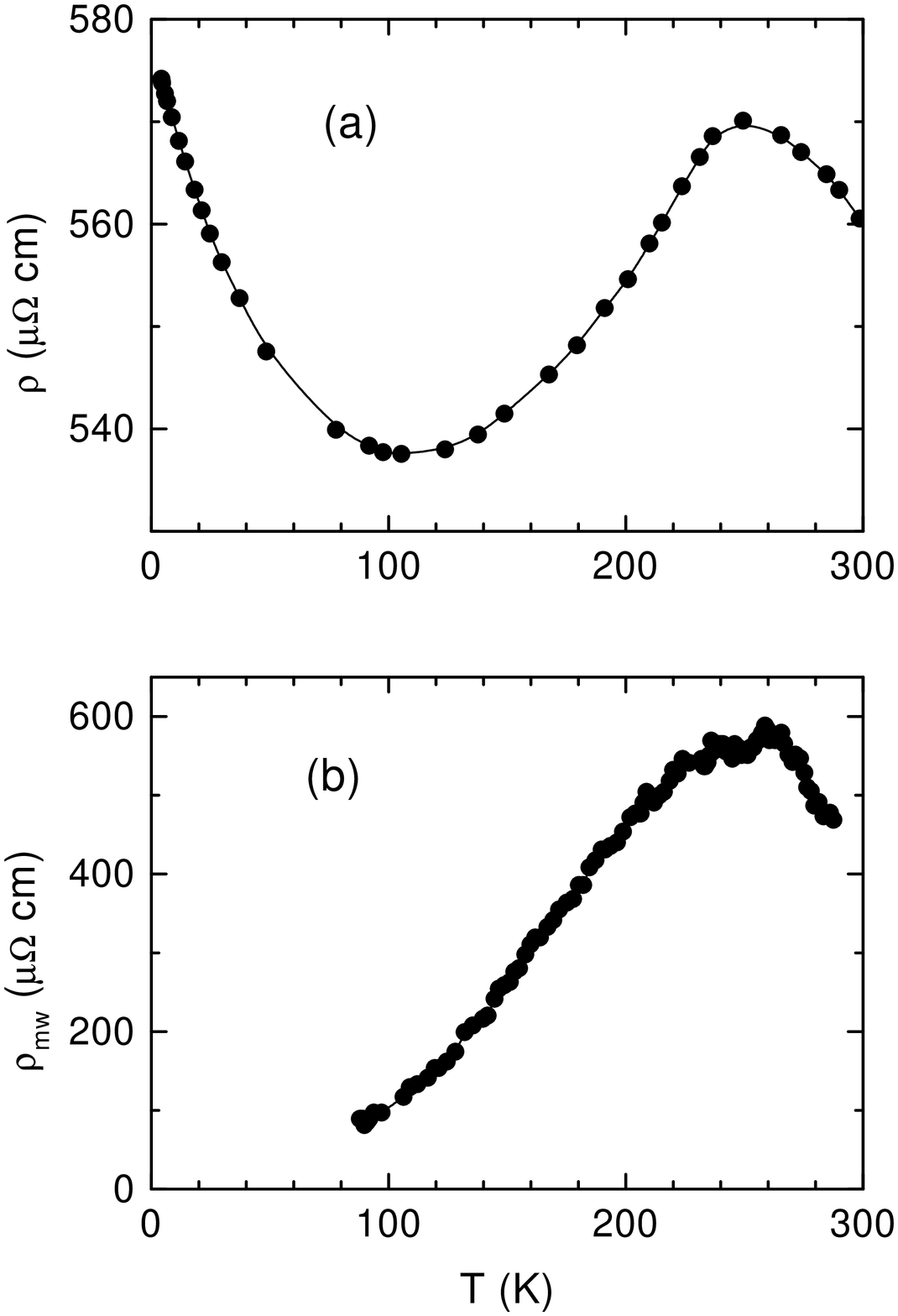,width=12cm}}
\vspace{15pt}
\centerline{Fig.2 to paper Belevtsev et al.}
\end{figure}

\begin{figure}
\centerline{\epsfig{file=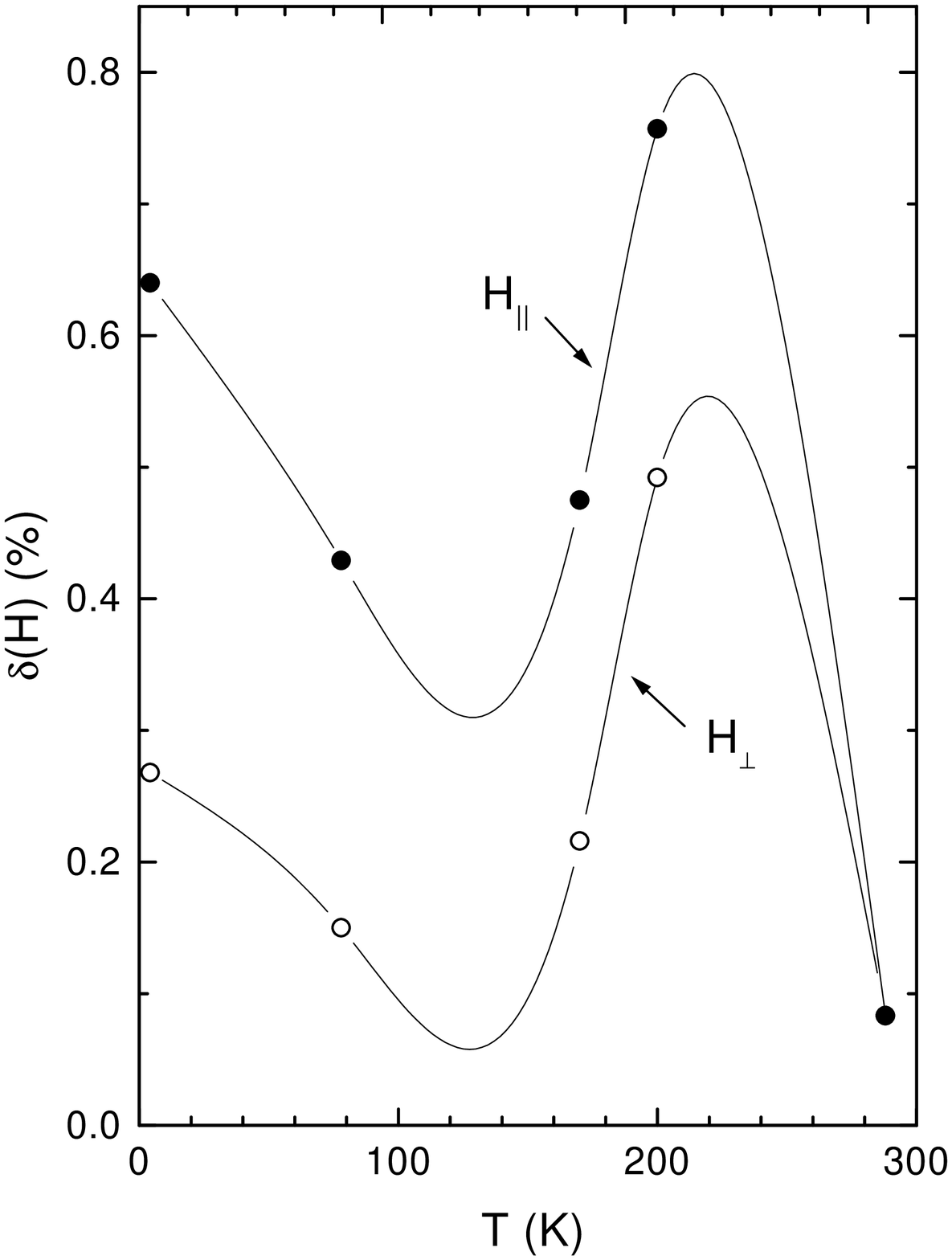,width=12cm}}
\vspace{15pt}
\centerline{Fig.3 to paper Belevtsev et al.}
\end{figure}

\end{document}